\documentclass[11pt]{article}

\usepackage[english]{babel}
\usepackage[latin1]{inputenc}
\usepackage{lmodern,csquotes}
\usepackage[T1]{fontenc}

\usepackage{amssymb,amsmath,amsthm,latexsym,amsfonts,amscd,dsfont,enumerate,bbm}
\usepackage{a4wide,graphicx,tikz}
\usepackage{xcolor} 
\graphicspath{ {.} }

\usepackage[backend=biber,style=authoryear-comp,maxcitenames=1,maxbibnames=99,sorting=nyt,sortcites=false]{biblatex}
\usepackage{pgfplots,pgfplotstable,booktabs,colortbl}
\pgfplotsset{compat=newest}

\pgfplotstableread{diff_skews_log_coords_n500_new_red2.txt}\logskew
\pgfplotstableread{diff_skews_log_coords_n20_new_red2.txt}\logskewtwenty

\definecolor{mylightyellow}{rgb}{1,1,.8}
\definecolor{mylightgreen}{rgb}{.8,1,.8}
\definecolor{mydarkred}{RGB}{178,34,34}
\definecolor{mydarkgreen}{RGB}{34,139,34}
\definecolor{mydarkblue}{RGB}{72,61,139}
\definecolor{mydarkyellow}{RGB}{218,165,32}

\usepackage[]{hyperref}
 \hypersetup{
 colorlinks=true,breaklinks=true,
 urlcolor= mydarkblue,linkcolor= mydarkblue,citecolor= mydarkgreen,
 pdfauthor={Dell'Acqua, E., Longoni, R., Pallavicini, A.},
}

\newtheorem{prop}{Proposition}[section]
\newtheorem{defn}{Definition}[section]

\newcommand{\Eq}[1]{{\small\begin{equation}{#1}\end{equation}}}

\newcommand{\Ex}[2]{\mathbb{E}_{#1}\!\left[\,#2\,\right]}

\newcommand{\ExC}[3]{\mathbb{E}_{#1}\!\left[\left.\,#2\,\right|\,#3\,\right]}

\newcommand{\onehalf}{\tfrac{1}{2}}

\title{Rough-Heston Local-Volatility Model}
\author{
Enrico Dall'Acqua\thanks{Politecnico Milano, {\tt enrico.dallacqua@mail.polimi.it}.}
\and
Riccardo Longoni\thanks{Intesa SanPaolo Milano, {\tt riccardo.longoni@intesasanpaolo.com}.}
\and
Andrea Pallavicini\thanks{Intesa SanPaolo Milano, {\tt andrea.pallavicini@intesasanpaolo.com}.}
}
\date{
\small First Version: October 22, 2021.  This version: \today
}

\begin{document}

\maketitle

\begin{abstract}

In industrial applications it is quite common to use stochastic volatility models driven by semi-martingale Markov volatility processes. However, in order to fit exactly market volatilities, these models are usually extended by adding a local volatility term. Here, we consider the case of singular Volterra processes, and we extend them by adding a local-volatility term to their Markov lift by preserving the stylized results implied by these models on plain-vanilla options. In particular, we focus on the rough-Heston model, and we analyze the small time asymptotics of its implied local-volatility function in order to provide a proper extrapolation scheme to be used in calibration.

\end{abstract}

\bigskip

\noindent {\bf AMS classification codes:} 65C05, 91G20, 91G60.

\newpage
{\small \tableofcontents}
\vfill
{\footnotesize \noindent The opinions here expressed  are solely those of the authors and do not represent in any way those of their employers.}
\newpage

\maketitle

\pagestyle{myheadings} \markboth{}{{\footnotesize E. Dall'Acqua, R. Longoni, A. Pallavicini, Rough-Heston Local-Volatility Model}}

\section{Introduction}
\label{sec:introduction}

Stochastic volatility (SV) models describe the price process of a financial asset by means of a continuous semi-martingale where the volatility coefficient is usually driven by a second semi-martingale Markov process. These models are successfully used in equity markets to price derivative options, see for instance \textcite{Heston1993}, but nevertheless they are unable to recover all the stylezed facts found in these markets. In particular, SV models are able to reproduce the leverage effect as described in \textcite{Hagan2002}, namely the correlation between spot price and volatility processes, but they seem to fail in describing the impact of past returns on the volatility, see \textcite{Rosenbaum2021} for a discussion. Moreover, when they are used for pricing, they are not capable of reproducing the correct form of the implied volatility skew at small times, as discussed in \textcite{Fukasawa2011,Fukasawa2019}.

In the last few years a new class of models, known as rough volatility (RV) models, has been proposed to recover these stylized facts. The idea behind RV models is a non-Markov process for the volatility with singular behaviour at small times. The origin of these models goes back to the paper of \textcite{Gatheral2018}, and since then two families of RV models have been developed and are currently studied, as categorized by \textcite{Bayer2020}. The first family consists of models where the volatility is a deterministic function of a fractional Brownian motion with Hurst parameter $H\in(0,\onehalf)$, see for instance the model proposed in \textcite{Gatheral2016}. The second family consists of models where the volatility process is described by a singular Volterra process, as the rough-Heston model introduced in \textcite{ElEuch2019}.

The main motivation of this paper is to understand how RV models can be used in practice in the activity of pricing and hedging derivatives, and in particular to address the problem of building a portfolio of plain-vanilla options that hedges against movements of the market implied volatility surface. This is a typical problem that arises with SV models but it is also shared with RV models. More precisely, sensitivities of the option price w.r.t.\ market volatilities, as computed by a SV model, may not give a good hedge against movements of market data which are not (typically) in line with what the model predicts. This problem has been solved for SV models by multiplying the volatility coefficient by a local-volatility term to obtain a local-stochastic volatility (LSV) model. This practice was first proposed in \textcite{Jex1999,Lipton2002} and allows, at least in principle, a perfect calibration to plain-vanilla options. We address also to \textcite{Ren2007,Guyon2012} for simple implementations of this idea that ensure a fast and reliable calibration.

In this contribution we discuss how to implement the same strategy with RV models, and in particular we focus on the rough-Heston model. Moreover, the discussion leads us to extend the recent asymptotic results of \textcite{DeMarco2022} and \textcite{Alos2022} on RV models with volatility driven by a fractional Brownian motion.

\medskip

The paper is organized as follows. In Section~\ref{sec:model} we discuss how to extend the rough-Heston model and we propose a new LSV model named HMLV mimicking the shape of the implied-volatility surface of the original model. Then, in Section~\ref{sec:numerics} we empirically verify the asymptotics property of the Rough-Heston model and we calibrate the HMLV model to plain-vanilla options.

\medskip

The authors thank Carlo Sgarra and Stefano De Marco for stimulating discussions on the subject of this article.

\section{The rough-Heston local-volatility model}
\label{sec:model}

In our analysis we focus on a specific RV model, the rough-Heston model first proposed in the work of \textcite{ElEuch2019}, which was subject of active research in the recent years. In this section we discuss how to extend this model to allow a perfect fit of plain-vanilla prices. We analyze two different proposals and we discuss how to preserve the shape of the implied volatility surface when extending the model.

\subsection{Extending the rough dynamics}
\label{sec:first}

The rough-Heston model is a generalization of the renowned Heston model first proposed in \textcite{Heston1993} where the variance process is replaced by a Volterra equation with a singular kernel as follows. We fix a filtered probability space $(\Omega, {\cal F}, {\mathbb P})$, with a filtration $({\cal F}_t)_{t\geq 0}$ representing the evolution of all the available information on the market. We assume that a risk-neutral measure $\mathbb Q$ equivalent to the physical measure $\mathbb P$ exists, and we introduce the following price dynamics under $\mathbb Q$.
\Eq{
dS^{\rm H}_t = S^{\rm H}_t \sqrt{v^{\rm H}_t} \,dB^{\rm H}_t \;,
\label{eq:rh}
}%
\Eq{
v^{\rm H}_t = v_0+\int_0^t K(t-u) \left( \lambda (\theta - v^{\rm H}_u) \,du + \nu \sqrt{v^{\rm H}_u} \,dW^{\rm H}_u \right) \;,
\label{eq:volterra}
}%
where $\lambda \ge 0$ is the speed of mean-reversion to the long-term level $\theta>0$, $\nu>0$ is the volatility of variance, $B^{\rm H}_t$ and $W^{\rm H}_t$ are two correlated Brownian motions under the risk-neutral measure with $d\langle B^{\rm H},W^{\rm H}\rangle_t=\rho\,dt$. The starting point $v_0$ of the variance process is a free parameter, while we assume that the spot price observed in the market is given by $S_0 := 1$. Then, we define the integration kernel $K$ as
\Eq{
K(t) := \frac{t^{H-\onehalf}}{\Gamma\!\left(H+\onehalf\right)}
\;,\quad
H \in \left(0,\onehalf\right] \;,
}%
where the Hurst parameter $H$ measures the roughness of the path, see \textcite{Gatheral2018}. More precisely the paths are locally H\"older-continuous of any order strictly less than $H$. We recall that the case $H=\onehalf$ corresponds to the case of a It\^o diffusion. In this case the model reduces to the standard Heston model. We assume a zero drift term for the stock price $S^{\rm H}_t$ for sake of exposition. We could add deterministic interest rates and dividends in a straightforward way. The existence of a strong solution of the above dynamics is shown in \textcite{ElEuch2019}.

The rough-Heston model can be calibrated to plain-vanilla options by using the characteristic function of the stock price which can be calculated by solving a fractional Riccati-Volterra equation, as shown in \textcite{ElEuch2019}.

As discussed in the introduction, we wish to build a hedging portfolio of plain-vanilla options against movements of the market implied-volatility surface, so that we need a set of parameters rich enough to perfectly fit the plain-vanilla prices in order to reproduce their price movements occurring in the market. With this in mind we investigate the following generalization of the rough-Heston model
\Eq{
d{\check S}^{\rm H}_t = {\check S}^{\rm H}_t \ell_{\rm H}(t,{\check S}^{\rm H}_t) \sqrt{v^{\rm H}_t} \,dB^{\rm H}_t \;,
\label{eq:rhlv}
}%
where the variance process is given by equation \eqref{eq:volterra}, and the diffusive coefficient of the stock price is enriched with a function $\ell_{\rm H}$ of both time and price. This is the standard approach adopted when introducing LSV models, see \textcite{Lipton2002} for details. In these models $\ell_{\rm H}$ is known as leverage function, a terminology we keep also in our case.

The leverage function could be calibrated to plain-vanilla options by following the same strategy proposed in \textcite{Guyon2012}, and refined in \textcite{Muguruza2019}, for LSV models. Here, we wish to highlight the issues arising from the specific form of the variance process we adopted for our model. The first step is to define the Markovian projection ${\hat S}^{\rm H}_t$ of the price process ${\check S}^{\rm H}_t$ as given by
\Eq{
d{\hat S}^{\rm H}_t = {\hat S}^{\rm H}_t {\hat \eta}_{\rm H}(t,{\hat S}^{\rm H}_t) \,d{\hat B}^{\rm H}_t \;,
\label{eq:lvH}
}%
where ${\hat B}^{\rm H}_t$ is a standard Brownian motion under the risk-neutral measure, and the local-volatility function ${\hat \eta}_{\rm H}$ is given by
\Eq{
{\hat \eta}_{\rm H}(t,x) := \ell_{\rm H}(t,x) \sqrt{ \ExC{0}{v^{\rm H}_t}{{\check S}^{\rm H}_t=x} } \;.
\label{eq:etaH}
}%
By \textcite{Gyongy1986} lemma we know that ${\hat S}^{\rm H}_t$ and ${\check S}^{\rm H}_t$ have the same marginal distributions and hence produce the same model prices of plain vanilla options. We remark that, as shown in \textcite{Brunick2013} the hypotheses of the lemma in the original paper can be extended to include continuous semi-martingales with measurable drift and volatility, as in our case. We refer also to \textcite{DeMarco2022} where this extension is used to define the implied local-volatility function of the rough-Bergomi model. 

Thus, calibration of model \eqref{eq:rhlv} consists in finding a leverage function such that plain-vanilla prices implied from the local-volatility dynamics \eqref{eq:lvH} with definition \eqref{eq:etaH} are equal to the corresponding market quotes. These plain-vanilla prices can be calculated by means of the Dupire formula described in \textcite{Dupire1994,Derman1994}. Particular care is needed to preserve some relevant properties of the original RV model, as we are discussing in the next sections.

We can summarize the above discussion by introducing the rough-Heston local-volatility model in the following definition.
\begin{defn}{\bf (Rough-Heston local-volatility model)}
\label{def:rhlvm}
The dynamics of the price process ${\check S}^{\rm H}_t$ under the rough-Heston local-volatility model is given by
\Eq{
d{\check S}^{\rm H}_t = {\check S}^{\rm H}_t {\hat \eta}_{\rm H}(t,{\check S}^{\rm H}_t) \sqrt{ \frac{v^{\rm H}_t}{\ExC{0}{v^{\rm H}_t}{{\check S}^{\rm H}_t}} } \,dB^{\rm H}_t \;,
}%
\Eq{
v^{\rm H}_t = v_0+\int_0^t \frac{(t-u)^{H-\onehalf}}{\Gamma\!\left(H+\onehalf\right)} \left( \lambda (\theta - v^{\rm H}_u) \,du + \nu \sqrt{v^{\rm H}_u} \,dW^{\rm H}_u \right) \;.
}%
\label{def:model}
\end{defn}%

The above equation is a rough differential equation of mean-field type. Only few papers in the literature analyze these equations, among them we cite the work of \textcite{Bailleul2020} which studies a simpler specification of the above equation to verify the existence of the solutions. In particular, they assume a more regular diffusive coefficient and a variance process given by a SDE driven by a fractional Brownian motion. The challenging extension of their work to our more general setting is outside the scope of our paper.

On the other hand, when we use the rough-Heston model in the practice, for instance to simulate the price process to calculate derivative prices, we follow the standard approach of approximating the model with a Markovian representation driven by semi-martingale processes, or Markovian lift, see \textcite{Jaber2019b}. Thus, we could work directly with this approximation to define a local-volatility extension. In the next section we recall the Markovian lift approximation and we show how to extend it with a local-volatility term.

\subsection{Extending the Markovian lift}
\label{sec:lift}

The rough-Heston model admits a representation in terms of an infinite dimensional stochastic-volatility process, known as Markovian lift and discussed in \textcite{Jaber2019a,Jaber2019b,Jaber2019c}. More explicitly, one writes the fractional kernel $K$ as a Laplace transform of a positive measure $\mu$ which is then approximated by a finite sum of Dirac masses
\Eq{
K(t) = \int_0^\infty e^{-\gamma t} \, \mu(d\gamma) \approx \sum_{i=1}^n c_i e^{-\gamma_i t}
}%
for a suitable choice of coefficients $\{c_i\}_{i=1:n}$ and $\{\gamma_i\}_{i=1:n}$ for an arbitrary large $n$. If we substitute this approximation in equation \eqref{eq:rh} we obtain a stochastic-volatility model with $n+1$ factors, a spot price process $S_t$ and $n$ driving factors $\{v^i_t\}_{i=1:n}$. The model can be written as
\Eq{
dS_t = S_t \sqrt{ v_t } \,dB_t \;,
\label{eq:lift}
}
whose variance process is given by
\Eq{
v_t := v_0 + \sum_{i=1}^n c_i \left( v^i_t + \lambda \theta \frac{1-e^{-\gamma_i t}}{\gamma_i} \right)
\label{eq:liftvar}
}%
with the driving processes $v^i_t$ defined as
\Eq{
dv^i_t = - (\gamma_i v^i_t + \lambda v_u) \,dt + \nu \sqrt{v_t} \,dW_t
}%
and starting at $v^i_0 = 0$, and the driving noises $B_t$ and $W_t$ are two correlated Brownian motions under the risk-neutral measure with $d\langle B,W\rangle_t = \rho\,dt$.

The existence of a strong solution for equation \eqref{eq:lift} and its convergence to the original rough-Heston model given by equation \eqref{eq:rh}, along with a criterium to select the location of the Dirac masses, are discussed in \textcite{Jaber2019b,Jaber2019c,Bayer2021}.

Now, we apply the machinery of LSV models presented in the previous section in order to obtain a LSV model which has as stochastic backbone the Markovian lift of the rough-Heston model. We name the resulting model as rough-Heston Markovian lift with local volatility, or simply HMLV.
\begin{defn}{\bf (HMLV model)}
The dynamics of the price process ${\check S}_t$ under the HMLV model is given by
\Eq{
d{\check S}_t = {\check S}_t {\hat \eta}(t,{\check S}_t) \sqrt{ \frac{v^{\phantom H}_t}{\ExC{0}{v_t}{{\check S}_t}} } \,dB_t \;,
}%
where $v_t$ is defined by equation \eqref{eq:liftvar}.
\label{def:approx}
\end{defn}%

The existence of a strong solution for LSV models, where the previous definition is an example, is an open problem discussed in the literature, see for instance \textcite{Lacker2020} for a review. Here, we address the reader also to the recent work of \textcite{Bayer2022} where the authors introduce a regularization scheme to assess the well-posedness of LSV models.

The above definition ensures that the HMLV model has the same marginal distribution as the local volatility model
\Eq{
d{\hat S}_t = {\hat S}_t {\hat \eta}(t,{\hat S}_t) \,d{\hat B}_t \;,
\label{def:approx_lv}
}%
where ${\hat B}_t$ is a standard Brownian motion under the risk-neutral measure. Thus, we can adopt the usual strategy to calibrate the local volatility model given by equation \eqref{def:approx_lv} to plain-vanilla prices to obtain the local-volatility function ${\hat \eta}$, which in turn ensures that also the HMLV model reproduces the same plain-vanilla prices. Yet, we recall that the market usually quotes plain vanilla options only on a selection of maturities and strikes, so that we need to introduce some interpolation and extrapolation schemes to properly define the local-volatility function, and, in doing so, we wish the HMLV model to preserve the asymptotic properties of the rough-Heston model. In particular, we wish to preserve the small-time behaviour of the implied volatility skew, as described in \textcite{Forde2021}.

In the next section we recall some asymptotic properties of the rough-Heston model and we derive the small-time behaviour of its implied local-volatility. In doing so we extend the results of \textcite{DeMarco2022} to the rough-Heston model. Then, we use this result to introduce an interpolation and extrapolation scheme for the local-volatility function which preserves this behaviour for the HMLV model.

\subsection{Local and implied volatilities in the rough-Heston model}
\label{sec:lv}

In general, if we are able to calculate plain-vanilla prices by means of a pricing model, such as the rough-Heston model of equation \eqref{eq:rh}, we can define the implied volatility ${\sigma}_{\rm H}$ as the function of time $t$ and strike $k$ given by inverting the following relationship for any time $t\ge\delta>0$.
\Eq{
{\rm Bl}(t,k,\sigma_{\rm H}(t,k)) = c(t,k) \;,
}%
where ${\rm Bl}(t,k,\sigma_{\rm H})$ is the Black pricing formula for call option with maturity $t$, strike $k$ and volatility $\sigma_{\rm H}$, and $c(t,k) := \Ex{0}{(S^{\rm H}_t - k)^+}$ is the model price of the option. We recall that we are assuming a zero drift term for the stock price for sake of exposition.

On the other hand, one can always find a local-volatility function $\eta_{\rm H}$ such that the call prices $c(t,k)$ satisfy the associated Dupire equation, see \textcite{Dupire1994,Derman1994}. Here, we write the Dupire equation with initial condition set at time $\delta>0$.
\Eq{
\left( \partial_t - \onehalf k^2 \eta_{\rm H}^2(t,k) \partial^2_k \right) c(t,k) = 0
\;,\quad
c(\delta,k) = \Ex{0}{(S^{\rm H}_\delta - k)^+} \;,
}%
which, by direct calculation, leads to the following relationship among the implied and the local volatilities.
\Eq{
\eta_{\rm H}^2
= \frac{
    \sigma_{\rm H}^2
  + 2 \sigma_{\rm H} t \,\partial_t \sigma_{\rm H}
}
{
  1
  + 2 k \sqrt{t}(y+ \sigma_{\rm H} \sqrt{t}) \partial_k \sigma_{\rm H}
  + k^2 \sigma_{\rm H} t \partial_k^2\sigma_{\rm H}
  + k^2 t y (y+\sigma_{\rm H} \sqrt{t}) (\partial_k \sigma_{\rm H})^2
}
\label{eq:eta}
}%
with $y := -\frac{1}{\sigma_{\rm H}\sqrt{t}}\log k -\onehalf\sigma_{\rm H} \sqrt{t}$ and $t\ge\delta>0$.

We make the choice of setting the initial condition at time $\delta>0$ since we wish to analyze the regularity of the implied and the local volatilities for small times, and the above equations may fail to exists at time $\delta=0$. Indeed, our plan is to use asymptotic results derived in \textcite{Forde2021} on $\sigma_{\rm H}$, which allow to expand the implied volatilities around time $t=0$. Then, we will use these results to study the limit of equation \eqref{eq:eta} for $\delta\rightarrow 0$.

In order to study the behaviour of the previous equation for small times, we need to know the regularity of $\sigma_{\rm H}$ and its derivatives. In doing so in \textcite{Forde2021} the authors consider a new coordinate system under which studying the small-time limit.
\Eq{
\theta := t
\;,\quad
\zeta := t^{H-\onehalf} \log k \;.
}%
We can also write the local and implied volatilities in terms of these new coordinates, and we get
\Eq{
{\bar\sigma}_{\rm H}(\theta,\zeta) := \sigma_{\rm H}\left(\theta,e^{\zeta \theta^{\onehalf-H}}\right)
\;,\quad
{\bar\eta}_{\rm H}(\theta,\zeta) := \eta_{\rm H}\left(\theta,e^{\zeta \theta^{\onehalf-H}}\right) \;.
}%
Then, we proceed by writing also equation \eqref{eq:eta} under these coordinates. The Jacobian of the coordinate transformation is given by
\Eq{
\begin{pmatrix}
\partial_\theta \\ \partial_\zeta
\end{pmatrix}
=
\begin{pmatrix}
\partial_t \theta & \partial_t \zeta \\ \partial_k \theta & \partial_k \zeta
\end{pmatrix}^{\!\!-1}
\begin{pmatrix}
\partial_t \\ \partial_k
\end{pmatrix} \;,
}%
so that we can write
\Eq{
\begin{pmatrix}
\partial_t \\ \partial_k
\end{pmatrix}
=
\begin{pmatrix}
\partial_\theta + (H-\onehalf) \theta^{-1} \zeta \,\partial_\zeta \\ \theta^{H-\onehalf} e^{- \zeta \theta^{\onehalf-H}} \,\partial_\zeta
\end{pmatrix} \;.
}%
The second derivative can be calculated by applying twice the first derivative
\Eq{
\partial^2_k = e^{- 2\zeta \theta^{\onehalf-H}} \left( \theta^{2H-1} \,\partial^2_\zeta - \theta^{H-\onehalf} \,\partial_\zeta \right) \;.
}%
We can substitute the derivatives and write for $0<\delta\le\theta$
\Eq{
{\bar\eta}_{\rm H}^2 = \frac{
  {\bar\sigma}_{\rm H}^2
  + 2 {\bar\sigma}_{\rm H} \theta \,\partial_\theta {\bar\sigma}_{\rm H}
  + {\bar\sigma}_{\rm H} (2H-1) \zeta \,\partial_\zeta {\bar\sigma}_{\rm H}
}
{
  1
  + 2 \theta^{H} (y - {\bar\sigma}_{\rm H} \theta^{\onehalf}) \,\partial_\zeta {\bar\sigma}_{\rm H}
  + {\bar\sigma}_{\rm H} \theta^{2H} \,\partial^2_\zeta {\bar\sigma}_{\rm H}
  + \theta^{2H} y (y+{\bar\sigma}_{\rm H} \theta^{\onehalf}) (\partial_\zeta {\bar\sigma}_{\rm H})^2
} \;.
}%

Before taking the small-time limit we have to analyze the regularity of the implied volatility. In the rough-Heston model the implied volatility ${\bar\sigma}_{\rm H}$ can be expanded for small values of $\theta$ at a given level $\zeta$ according to equation (44) in \textcite{Forde2021}.
\begin{equation}
{\bar\sigma}_{\rm H}(\theta,\zeta) = \alpha_0(\zeta) + \alpha_1(\zeta) \theta^{2H} + o(\theta^{2H})
\end{equation}
for $\theta\ge 0$ and some functions $\alpha_0(\zeta)$ and $\alpha_1(\zeta)$, so that we obtain in the small-time limit
\begin{equation}
\theta \,\partial_\theta {\bar\sigma}_{\rm H}(\theta,\zeta) = 2H \alpha_1(\zeta) \theta^{2H} + o(\theta^{2H}) \;.
\end{equation}
Moreover, according to equation (32) in \textcite{Forde2021} we can expand the leading term of the previous time expansion around $\zeta=0$, so that we can assume the existence of the implied volatility and its derivatives w.r.t.\ $\zeta$ in the small-time limit. Thus, we can write for $\theta = \delta$ and for small values of $\delta$.
\begin{equation}
{\bar\eta}_{\rm H}(\delta,\zeta) = {\bar\sigma}_{\rm H}(\delta,\zeta)
\,\frac{
  \sqrt{ 1 - (1-2H) \zeta \,\partial_\zeta \log {\bar\sigma}_{\rm H}(\delta,\zeta) }
}
{
  1 - \zeta \,\partial_\zeta \log {\bar\sigma}_{\rm H}(\delta,\zeta)
}
+ o(1) \;.
\end{equation}%

Then, since the right-hand side of the above equation is well defined for any value of $\delta\ge 0$, we can take the small-time limit and we obtain by direct inspection for the at-the-money level the following result
\begin{equation}
{\bar\eta}_{\rm H}(0,0) = {\bar\sigma}_{\rm H}(0,0) \;.
\end{equation}
Moreover, we can expand both the local and the implied volatility around $\zeta=0$ to get the relationship between the skews
\begin{equation}
\partial_\zeta {\bar\eta}_{\rm H}(0,0) = \left(H+\frac{3}{2}\right) \,\partial_\zeta {\bar\sigma}_{\rm H}(0,0) \;,
\end{equation}%
which we can summarize in the following proposition in term of the original functions $\sigma_{\rm H}$ and $\eta_{\rm H}$.
\begin{prop} {\bf (``H+3/2'' rule)}
The skews of the local and implied volatilities in the rough-Heston model satisfy the small-time limit
\Eq{
\lim_{\zeta\rightarrow 0}\lim_{\theta\rightarrow 0}
\frac{\eta_{\rm H}(\theta,k(\theta,\zeta)) - \eta_{\rm H}(\theta,-k(\theta,\zeta))}
     {\sigma_{\rm H}(\theta,k(\theta,\zeta)) - \sigma_{\rm H}(\theta,-k(\theta,\zeta))}
=
H + \frac{3}{2} \;,
}%
where we define the strike $k(\theta,\zeta) := e^{\zeta \theta^{\onehalf-H}}$.
\label{prop:32rule}
\end{prop}

The result given by proposition~\ref{prop:32rule} extends the results of \textcite{DeMarco2022} to the case of the rough-Heston model by confirming the same skew behaviour also for this model and by highlighting that it depends only on the regularity of the implied volatility in the small-time limit.

\subsection{Interpolation scheme for the HMLV model}
\label{sec:interpolation_scheme}

In view of the results of the previous section, we wish to define an interpolation and extrapolation scheme for the local-volatility ${\hat \eta}$ to be used in the HMLV model that preserves the small-time property of the rough-Heston model. We assume that the market is quoting plain-vanilla prices with maturities $\{T_i\}_{i=1:N}$ and for each $T_j$ with strikes $\{K_{ij}\}_{j=1:M}$. 

We define the local volatility by means of an interpolation and extrapolation function $\psi$ in the coordinates $\theta$ and $\zeta$ and based on the knowledge of the local-volatility on the same set of times and strikes where plain-vanilla options are quoted. For $t>\delta$ we write 
\Eq{
{\hat \eta}(t,k) := \psi\left(t,t^{H-\onehalf}\log k;\{{\hat \eta}_{ij}\}\right)
\;,\quad
{\hat \eta}_{ij} = {\hat \eta}(T_i,K_{ij}) \;,
}%
while for $t<\delta$ we extrapolate flat in time along the strike coordinate. In the numerical investigations we choose $\delta$ to be smaller than the first step in the discretization of equation \eqref{def:approx}. In particular, we select $\psi(\theta,\zeta)$ being a piece-wise constant function of $\theta$, while along $\zeta$ we select a cubic monotone spline for interpolation and a constant function for extrapolation. This choice can be viewed as a generalization of what is done in the work of \textcite{Nastasi2020}.

\section{Numerical investigations}
\label{sec:numerics}

In this section we perform two numerical investigations. First, we check empirically the asymptotic behaviour established in proposition~\ref{prop:32rule} for the local and implied volatilities of the rough Heston model. Then, we verify that the HMLV model is able to reprice the plain-vanilla option prices quoted in the market. In our analysis we use the same rough-Heston parameters used by \textcite{Jaber2019c}, namely $v_0=0.02$, $\theta=0.02$, $\lambda=0.3$, $\nu=0.3$, $\rho=-0.7$ and Hurst index $H=0.1$.

\subsection{Empirical evidence of the rough-Heston behaviour}
\label{sec:empirical}

In order to empirically verify the asymptotic behaviour of the local and implied volatility skews of the rough-Heston model, we simulate $M = 1\times10^5$ Monte Carlo samples of its Markovian lift, as given in equation~\eqref{eq:lift}. In the paper of \textcite{Jaber2019c} an empirical analysis of the Markovian lift is performed, and it is shown that a number $n = 20$ of driving factors for the volatility process~\eqref{eq:liftvar} is enough to generate implied-volatility smiles that are close to those of the original rough-Heston model for maturities ranging from one week to more than one year.

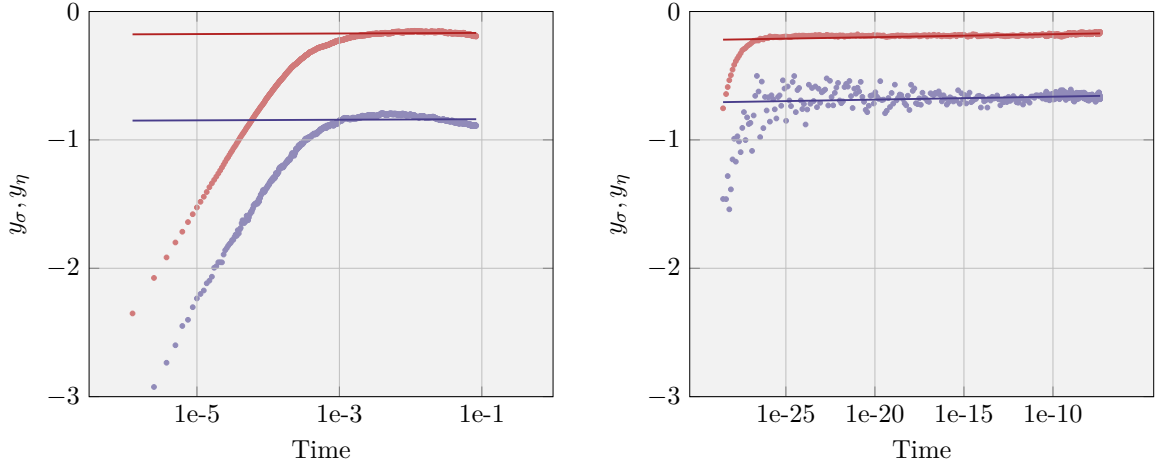
\begin{figure}
\begin{center}
\scalebox{0.9}{%
\begin{tikzpicture}
\begin{axis}[xlabel=Time,
             ylabel={$y_\sigma,y_\eta$},
             ylabel style={overlay},
             xmin=-15, xmax=-0,
             xtick={-16.11809565,-11.51292546,-6.907755279,-2.302585093},
             xticklabels={1e-7,1e-5,1e-3,1e-1},
             ymin=-3, ymax=0,
             grid=major,
             legend style={legend pos=south west},
             axis background/.style={fill=gray!10},
             set layers,mark layer=axis background]
\addplot [color=mydarkblue!60,only marks,mark=*,mark size=1pt] table [y=diff_log_skews_bs,x=log_maturities] from \logskewtwenty;
\addplot [color=mydarkblue,thick,smooth] table [y=diff_log_skews_bs_fit,x=log_maturities] from \logskewtwenty;
\addplot [color=mydarkred!60,only marks,mark=*,mark size=1pt] table [y=diff_log_skews_lv,x=log_maturities] from \logskewtwenty;
\addplot [color=mydarkred,thick,smooth] table [y=diff_log_skews_lv_fit,x=log_maturities] from \logskewtwenty;
\end{axis}
\end{tikzpicture}
\hspace*{1cm}
\begin{tikzpicture}
\begin{axis}[xlabel=Time,
             ylabel={$y_\sigma,y_\eta$},
             ylabel style={overlay},
             xmin=-70, xmax=-10,
             xtick={-57.56462732,-46.05170186,-34.53877639,-23.02585093},
             xticklabels={1e-25,1e-20,1e-15,1e-10},
             ymin=-3, ymax=0,
             grid=major,
             legend style={legend pos=south west},
             axis background/.style={fill=gray!10},
             set layers,mark layer=axis background]
\addplot [color=mydarkblue!60,only marks,mark=*,mark size=1pt] table [y=diff_log_skews_bs,x=log_maturities] from \logskew;
\addplot [color=mydarkblue,thick,smooth] table [y=diff_log_skews_bs_fit,x=log_maturities] from \logskew;
\addplot [color=mydarkred!60,only marks,mark=*,mark size=1pt] table [y=diff_log_skews_lv,x=log_maturities] from \logskew;
\addplot [color=mydarkred,thick,smooth] table [y=diff_log_skews_lv_fit,x=log_maturities] from \logskew;
\end{axis}
\end{tikzpicture}}
\end{center}
\caption{Differences between logarithms of at-the-money implied-volatility skews (in blue) or of local-volatility skews (in red) and the theoretical level $\beta_{\rm H}$ as a function of time, see equation~\eqref{eq:fit} for details. Left panel $n=20$ driving factors, right panel $n=500$. Points are empirical estimates, continuous lines are linear regressions discarding estimates below a critical time.}
\label{fig:skews_log_coords}
\end{figure}

If we consider this setting, we obtain a set of mean-reversion speeds $\gamma_i$ ranging from $1.76\times10^{-4}$ to $6.42\times10^{3}$, which in turn implies that the shortest time-scale modelled by the Markovian lift is about $\tau^{20}_{\rm short} :=1.56\times10^{-4}$ years. As a consequence, we can probe the scaling rule described by proposition~\ref{prop:32rule} only for longer time-scales. In the left panel of figure~\ref{fig:skews_log_coords} we show on the $y$-axis the behaviour of implied and local volatility skews as given by
\Eq{
y_\sigma := \ln\left(-\partial_k{\sigma}_{\rm H}(t,1)\right) - \beta_{\rm H} \ln t
\;,\quad
y_\eta := \ln\left(-\partial_k{\eta}_{\rm H}(t,1)\right) - \beta_{\rm H} \ln t
\;,
\label{eq:fit}
}%
where $\beta_{\rm H} := H - \frac{1}{2}$ is the theoretical slope obtained from the implied-volatility expansion of \textcite{Forde2021}, and it is equal to $-0.4$ in our setting where $H=0.1$. Indeed, we can see that the estimators are reliable only after $\tau^{20}_{\rm short}$, when their distribution flattens around a horizontal line. In particular, we can consider a linear regression of the the skews of implied (and local) volatilities with estimates after a critical time $T_\sigma^{20}$ (and $T_\eta^{20}$) greater than $\tau^{20}_{\rm short}$, and we fix it so that the regression slope is equal to $\beta_{\rm H}$. The resulting regression intercepts $\alpha_\sigma$ and $\alpha_\eta$ allow to estimate the scaling rule of the skew ratio given by proposition~\ref{prop:32rule} as given by $\exp\{\alpha_\eta - \alpha_\sigma\}$. In this setting we obtain an estimate of the skews ratio equal to $1.96$, with the critical time scales $T_\sigma^{20}=4.60\times10^{-4}$ and $T_\eta^{20}=1.72\times10^{-3}$, to be compared to the value of $1.60$ given by the proposition. The skew ratio with the setting $n=20$ seems far from the theoretical results and close to the classical result of $2.00$ derived in \textcite{Berestycki2004} for non-rough models, although the skew slopes show an agreement with the theory after removing estimates below critical times.

In order to explore a smaller time regime and to improve the accuracy of our investigation, we increase the number of driving factors to $n = 500$, and we obtain that the mean-reversion speeds range from $3.74\times10^{-29}$ to $2.70\times10^{28}$, while the smallest time scale is about $\tau^{500}_{\rm short} := 3.71\times10^{-29}$ years. We repeat the exercise just performed for the case $n=20$ also for the case $n=500$, and we show the results in the right panel of figure~\ref{fig:skews_log_coords}. We notice that in this second case that the estimates are noisier, but their allignment around a horizontal lines shows a lesser bias. In this setting we obtain an estimate of the skews ratio equal to $1.63$, with the critical time scales $T_\sigma^{500}=5.72\times10^{-28}$ and $T_\eta^{500}=1.71\times10^{-28}$, to be compared to the value of $1.60$ given by the proposition. Now, the empirical results align with the theory.

\subsection{Calibration to plain-vanilla options}
\label{sec:calibration}

\begin{figure}
\includegraphics[width=\textwidth]{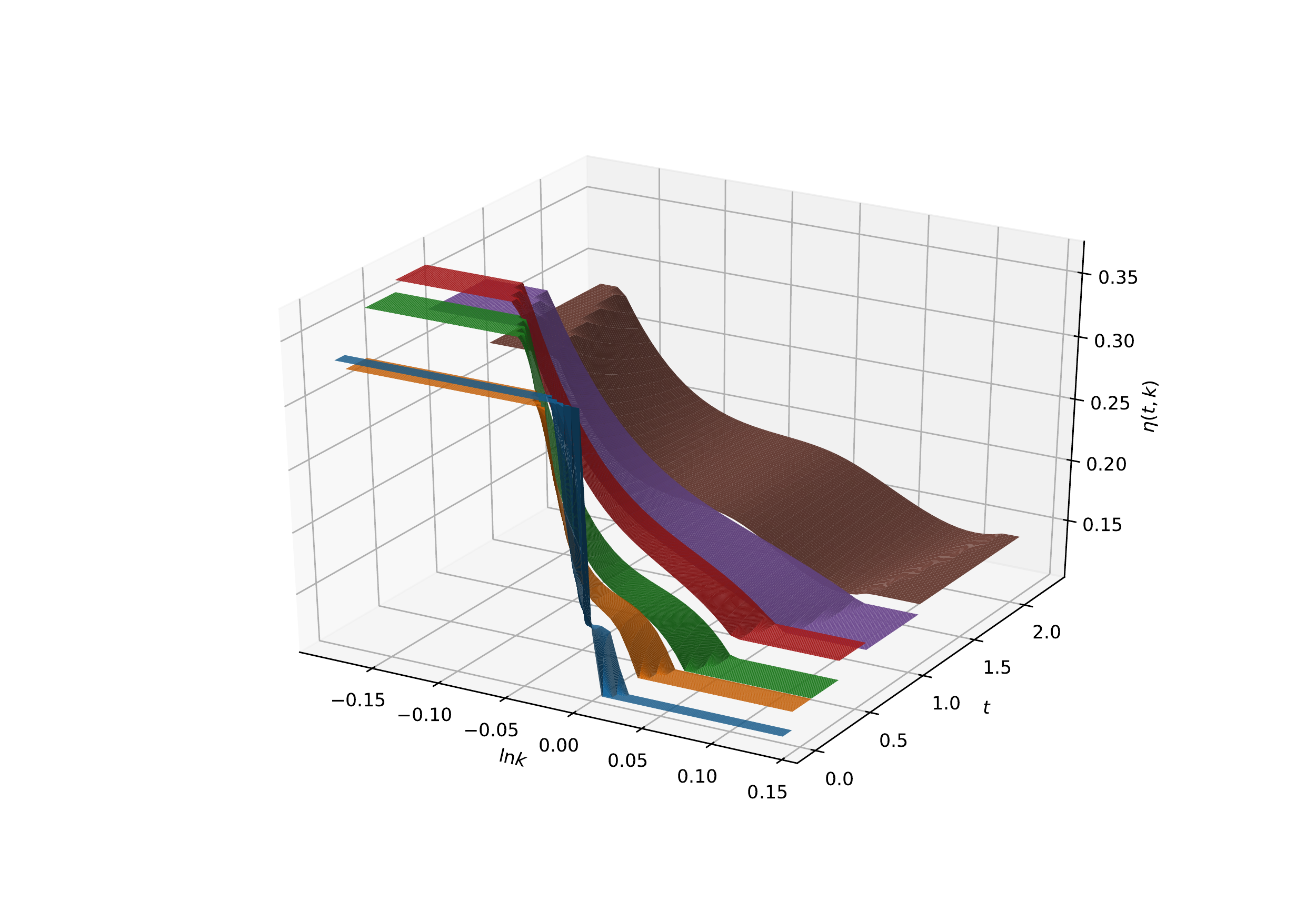}
\centering
\caption{The local-volatility function calibrated on market quotes of plain-vanilla options on Dow Jones Euro STOXX 50 as of 15 September 2021. On the left axis the logarithm of the moneyness, on the right axis the option maturity.}
\label{fig:lv_surf}
\end{figure}

\begin{table}
\begin{center}
\pgfplotstabletypeset[
    row sep=\\,
    col sep=&,
    columns={k1,e1,k2,e2,k3,e3,k4,e4,k5,e5,k6,e6},
    every even column/.style={column name=$k$, fixed, fixed zerofill, precision=2, dec sep align={|c}, clear infinite, column type/.add={|}{}},
    every first column/.style={column name=$k$, fixed, fixed zerofill, precision=2, dec sep align, clear infinite},
    every odd column/.style={column name=$\epsilon$, fixed, fixed zerofill, precision=2, dec sep align, clear infinite},
    every head row/.style={before row={\toprule \multicolumn{4}{c}{1m} & \multicolumn{4}{|c}{3m} & \multicolumn{4}{|c}{6m} & \multicolumn{4}{|c}{9m} & \multicolumn{4}{|c}{15m} & \multicolumn{4}{|c}{27m} \\},after row=\midrule},
    every last row/.style={after row=\bottomrule},
]{
k1 & e1 & k2 & e2 & k3 & e3 & k4 & e4 & k5 & e5 & k6 & e6\\
0.97 & -0.58 & 0.94 & -1.75 & 0.92 & -2.25 & 0.9 & -0.06 & 0.87 & -0.31 & 0.85 & 0.13\\
0.99 & 0.98 & 0.97 & 0.02 & 0.97 & 0.14 & 0.96 & 2.55 & 0.94 & 3.04 & 0.93 & 2.28\\
1 & 1.08 & 1 & 0.81 & 1 & 1.87 & 1 & 3.56 & 1 & 0.6 & 1 & 2.99\\
1.01 & 1.98 & 1.02 & 1.17 & 1.03 & 2.56 & 1.03 & 1.35 & 1.05 & 1.91 & 1.07 & 4.32\\
1.02 & 2.71 & 1.04 & 1.53 & 1.06 & 3.3 & 1.07 & 2.51 & 1.1 & 3.06 & 1.13 & 4.89\\
}
\end{center}
\caption{Re-pricing errors for plain-vanilla options used in the calibration phase expressed as differences between market and model implied-volatilities. For each expiry date the table shows the option moneyness $k$ and the error $\epsilon$ on volatility.}
\label{table:errors_bps}
\end{table}

We consider plain-vanilla options quoted in the market on 15 September 2021 with Dow Jones Euro STOXX 50 as underlying asset\footnote{Dow Jones Euro STOXX 50 is an equity index resulting from a weighted (on market capitalization) average of the values of the 50 most important companies in the Eurozone.}. We calibrate a local-volatility surface (shown in figure~\ref{fig:lv_surf}) as discussed in section~\ref{sec:interpolation_scheme} on a subset of quotes chosen among the most liquid and with the first expiry being one month afar. More precisely, a grid of market strikes is associated to each market maturity in order to form a lattice over which the local volatility surface is parametrized. Following the iterative approach presented in \textcite{Nastasi2020}, the initial nodal values (which correspond to the market implied volatilities) are updated until adherence to market prices by means of the implied volatilities associated to the local volatility surface undergoing calibration. What is innovative in our approach is that, we recall, between market maturities the interpolator is a constant function of time in the $\zeta$ coordinate.

We build an instance of the HMLV model upon the calibrated local-volatility surface by selecting as in \textcite{Jaber2019c} $n=20$ driving factors for the volatility process. We simulate $M=1\times10^4$ sample paths for the HMLV model and we re-price the options used in the calibration phase. The leverage function is calibrated by means of the same Monte Carlo simulation used for pricing by following the approach discussed in \textcite{Muguruza2019}. In table~\ref{table:errors_bps} we report the differences in basis points between the market volatilities and the model-implied ones. We get a maximum error below five basis points.

\section{Conclusions and future developments}
\label{sec:conclusion}

In this paper we presented a strategy to extend a RV model to include a local-volatility term in the same spirit of LSV models. In particular, we focused on the rough-Heston model and, after having discussed the difficulties arising in a straightforward extension, we introduced the HMLV model, a LSV model based on the Markovian lift of the rough-Heston model, and we discussed how to define the local-volatility function to preserve the asymptotic properties of the originating RV model. In doing so we extended the results of \textcite{DeMarco2022} on the small-time behaviour of the ratio between the skews of the local and the implied volatilities.

We leave for a future development the numerical investigation of exotic option pricing with the HMLV model, the empirical analysis of the samll-time behaviour of the skews with different simulation schemes, such as the one recently proposed in \textcite{Gatheral2021}, and the more ambitious goal of considering the local-volatility extension of the rough-Heston model without using the Markovian lift.

\printbibliography

\end{document}